\documentstyle[twocolumn,eqsecnum,prl,aps,psfig]{revtex}  
\begin{document}
\draft
\title{Non-Gaussian fixed point candidates \\ in the 4D compact U(1) gauge
theories}
\author{W. Franzki, J.~Jers{\'a}k\footnotemark[1]}
\address{Institut f{\"u}r
Theoretische Physik E, RWTH Aachen, Germany},
\author{C.~B.~Lang}
\address{Institut f{\"u}r Theoretische Physik,
Karl-Franzens-Universit\"at Graz, Austria}
\author{T.~Neuhaus}
\address{FB8 Physik, BUGH Wuppertal, Germany}
\date{\today}
\maketitle
\begin{abstract}
  Some interesting nonperturbative properties of the strongly coupled
  4D compact U(1) lattice gauge theories, both without and with matter
  fields, are pointed out. We demonstrate that the pure gauge theory
  has a non-Gaussian fixed point with $\nu = 0.365(8)$ at the second
  order confinement-Coulomb phase transition. Thus a non-asymptotic
  free and nontrivial continuum limit of this theory, and of its
  various dual equivalents, in particular of a special case of the
  effective string theory, can be constructed. Including a scalar
  matter field (compact scalar QED), we confirm the Gaussian behavior
  at the endpoint of the Higgs phase transition line. In the theory
  with both scalar and fermion matter fields, we demonstrate the
  existence of a tricritical point. Here, the chiral symmetry is
  broken, and the mass of unconfined composite fermions is generated
  dynamically. Appart from the Goldstone bosons, the spectrum contains
  also a massive scalar.  This resembles the Higgs-Yukawa sector of
  the SM, albeit of dynamical origin, like the Nambu--Jona-Lasinio
  model.  However, the scaling behavior is different from that in the
  NJL model and the nonperturbative renormalizability might thus be
  possible.
\end{abstract}
\narrowtext

\section{Introduction}

The only firmly established quantum field theories in four dimensions
(4D) are either asymptotically free or so-called trivial theories.
Both are defined in the vicinity of Gaussian fixed points. But in
abelian gauge theories at strong coupling several candidates for a
non-Gaussian fixed point exist. The best known example is the
noncompact QED with a chiral phase transition. In this contribution we
describe the recent progress in a systematic study of critical
behavior in several compact U(1) gauge models on the lattice, both
pure and with matter fields. It provides us with an information about
the possible fixed point structure of compact QED. (Various results in
this report have been obtained by various subsets of the present
authors.)

\footnotetext[1]{The speaker at the Warsaw conference}
\section{Pure gauge theory}

\subsection{Problem of the two-state signal}

First we reconsider the oldest candidate for a non-Gaussian fixed
point in the 4D lattice field theory, the phase transition between the
confinement and the Coulomb phases in the pure compact U(1) gauge
theory with Wilson action and extended Wilson action,
\begin{equation}\label{ACTION}
         S = -\sum_P w_P
              \left [\beta \cos(\Theta_P) + \gamma
                \cos(2\Theta_P)\right ].
\end{equation}
Here  $w_P = 1$ and $\Theta_P \in [0,2\pi)$ is the plaquette angle,
i.e. the argument of the product of U(1) link variables along a
plaquette $P$. Taking $\Theta_P = a^2gF_{\mu\nu}$, where $a$ is the
lattice spacing, and $\beta + 4\gamma = 1/g^2$, one obtains for weak
coupling $g$ the usual continuum action $S =\frac{1}{4} \int
d^4xF_{\mu\nu}^2$.

The detailed investigations performed as usual on toroidal lattices were
hindered mainly by a weak two-state signal \cite{JeNe83} at $\gamma =
0$.  This could either be a finite size effect, or it could imply that
the phase transition at $\gamma = 0$ is actually of $1^{\rm st}$
order, preventing a continuum limit there.

In the model with the extended Wilson action (\ref{ACTION}), it was
found that the confinement-Coulomb phase transition is clearly of
$1^{\rm st}$ order for $\gamma \ge 0.2$, and weakens with decreasing
$\gamma$. Various studies trying to take into account finite size
effects suggested that the transition becomes $2^{\rm nd}$ order at
slightly negative $\gamma$ \cite{EvJe85}, or around $\gamma =
0$\cite{La86,La87b}.  Nevertheless, the two-state signal persisted on finite
size lattices even at $\gamma = -0.5$ \cite{EvJe85}.

We demonstrate that the problems encountered, when considering the
continuum limit at this phase transition, can be surmounted. The clues
are the observation that the two-state signal disappears on lattices
with sphere-like topology, the construction of homogeneous spherical
lattices, the use of modern finite size scaling (FSS) analysis
techniques, and larger computer resources. A more detailed account of
our work, as well as relevant references, can be found in
Refs.~\cite{JeLa96a,JeLa96b}.

\subsection{Spherical lattice}
Two of the present authors performed simulations at $\gamma = 0$,
using the 4D surface of a 5D cubic lattice instead of the torus. They
observed that on such lattices with sphere-like topology the two-state
signal vanishes \cite{LaNe94a,LaNe94b}. This suggests that the two-state
signal at $\gamma \le 0$ is related to the nontrivial topology of the
toroidal lattice.  It has been checked in spin models, that weak
two-state signals are not washed out on lattices with sphere-like
topology, if they are due to a genuine $1^{\rm st}$ order
transition\cite{HoJa96}.  However, the lattice on the surface of a
cube is rather inhomogeneous and causes complex finite size effects,
preventing a reliable FSS analysis.

For our present study at $\gamma \le 0$, we have chosen lattices with
sphere-like topology, but make them approximately homogeneous. We 
use lattices obtained by projecting the 4D surface $S\!H\,[N]$ of a
5D cubic lattice $N^5$ onto a concentric 4D sphere.  On such a
spherical lattice $S\,[N]$, the curvatures concentrated on the
corners, edges, etc., of the original lattice $S\!H\,[N]$ are
approximately homogenized over the whole sphere by the weight factors
\begin{equation}
             w_P = A'_P/A_P  \label{WP}
\end{equation}
in the action (\ref{ACTION}). $A_P$ and $A'_P$ are the areas of the
plaquette $P$ on $S\,[N]$, and of its dual, as on any irregular, e.g.
random lattice \cite{ChFr82}.  

It has been checked in some spin and gauge models with $2^{\rm nd}$
order transitions that universality for spherical lattices holds, and
that the FSS analysis works very well if $V^{1/D}$ is used as a linear
size parameter, $V\equiv\frac{1}{6}\sum_P w_P $ being the volume of
the sphere $S\,[N]$ \cite{HoJa96,HoLa96}.

\subsection{$2^{\rm nd}$ order scaling behavior}
The measurements have been performed at $\gamma = 0, -0.2, -0.5$ on
lattices of the sizes $N = 4, ..., 12$.  We find that at the
confinement-Coulomb phase transition at strong bare gauge coupling,
$g$ = O(1), the model exhibits a $2^{\rm nd}$ order scaling behavior
well described by the values of the correlation length critical
exponent $\nu$ in the range $\nu = 0.35 - 0.40$. The FSS behavior of
the Fisher zero, specific heat, some cumulants and pseudocritical
temperatures has been studied.

\begin{figure}[tbp]
  \begin{center}
    \leavevmode
     \psfig{file=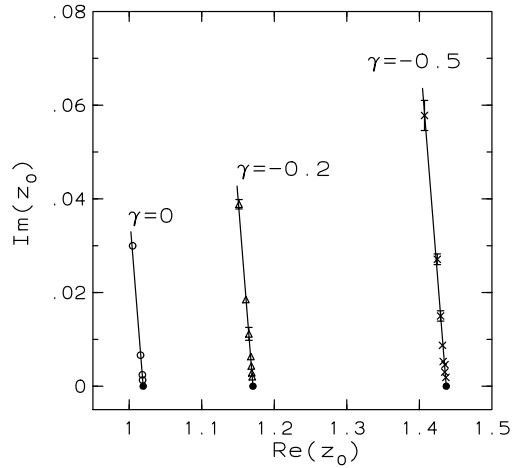,width=7.8cm,angle=180}
       \vspace{-1.7cm}
   \caption{Paths of the Fisher zeros $z_0$ in the complex $\beta$ plane. }
    \label{fig:1}
  \end{center}
\end{figure}

The most reliable measurement is provided by the FSS analysis of the
first zero $z_0$ of the partition function in the complex plane of the
coupling $\beta$ (Fisher zero). Applying the multihistogram
reweighting method for its determination, we have studied the approach
of $z_0$ to the real axis, shown in Fig.~\ref{fig:1}. The expected FSS
behavior
\begin{equation}
          \mbox{Im}\,z_0 \propto V^{-1/D\nu}    \label{IMZ}
\end{equation}
has been used for measuring $\nu$. This has turned out to be superior
to -- though consistent with -- the more common FSS analysis of
specific heat and cumulant extrema.  

\vspace{-2cm}

\begin{figure}[tbp]
  \begin{center}
    \leavevmode
     \psfig{file=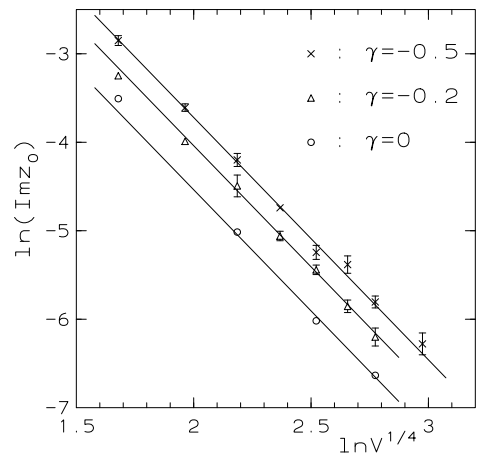,width=7.8cm,angle=180}
     \nopagebreak
     \vspace*{-1.7cm}
   \caption{ Joint fit to the FSS of $\mbox{Im}\,z_0$. }
    \label{fig:2}
  \end{center}
\end{figure}

\begin{figure}[tbp]
  \begin{center}
    \leavevmode
     \psfig{file=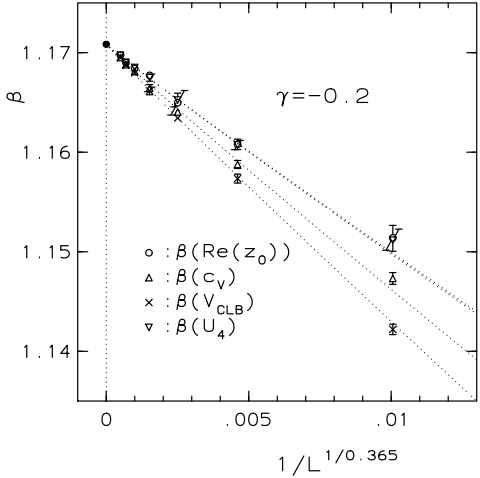,width=7.8cm,angle=180}
       \vspace{-1.7cm}
   \caption{Consistency of the FSS of various pseudocritical
     temperatures with the value $\nu = 0.365$. }
    \label{fig:3}
  \end{center}
\end{figure}

The joint fit to the data at all three $\gamma$ values, shown in
Fig.~\ref{fig:2}, gives
\begin{equation}
         \label{NU}       \nu = 0.365(8) \;.
\end{equation}
In Fig.~\ref{fig:3} we show that the scaling behavior of the
pseudocritical temperatures, which have been determined from several
different observables, is consistent with the value (\ref{NU}).
Further data and technical details are presented in
Ref.~\cite{JeLa96b}.

\vspace{-2cm}


\vspace{-2cm}


\subsection{Physical implications}

The results for $\nu$ are quite different from $\nu=0.25,$ expected at
a $1^{\rm st}$ order transition, as well as from $\nu = 0.5,$ obtained
in a Gaussian theory or in the mean field approximation. This strongly
suggests the existence of a continuum pure U(1) gauge theory with
properties different from theories governed by Gaussian fixed points
with or without logarithmic corrections. It can be obtained from the
lattice theory by the RG techniques. To our knowledge, the existence
of such a continuum quantum field theory in 4D is in no way indicated
by the perturbation expansion.

The physical content of the continuum limit of the pure compact U(1)
gauge theory at the confinement-Coulomb phase transition depends on
the phase from which the critical line is approached. In the
confinement phase, a confining theory with monopole condensate is
expected, as the string tension scales with a critical exponent
consistent with the value (\ref{NU})\cite{JeNe85}.  The physical
spectrum consists of various gauge balls, whose spectrum is currently
under investigation \cite{CoFr97a}. In the Coulomb phase, massless photon
and massive magnetic monopoles should be present. The renormalized
electric charge $e_r$ is large but finite\cite{JeNe85}, and has
presumably a universal value. The numerical result $e_r^2/4\pi =
0.20(2)$\cite{JeNe85} agrees with the L\"uscher bound.

Since the pure U(1) lattice gauge theory with the Villain (periodic
Gaussian) action presumably belongs to the same universality class,
rigorous dual relationships imply that also the following 4D models
possess a continuum limit described by the same fixed point: the
Coulomb gas of monopole loops\cite{BaMy77}, the noncompact U(1) Higgs
model at large negative squared bare mass (frozen 4D superconductor)
\cite{Pe78,FrMa86}, and an effective string theory equivalent to this
Higgs model\cite{PoSt91,PoWi93}.

These findings raise once again the question, whether in strongly
interacting 4D gauge field theories further non-Gaussian fixed points
exist, that might possibly be of interest for theories beyond the standard
model. 

The pursuit of this question requires an introduction of matter
fields. Therefore we investigate, what extensions of the pure compact
U(1) gauge theory might hide interesting fixed points. We introduce
fermion and scalar matter fields of unit charge, either each separately or
both simultaneously.

\section{Status of compact QED with staggered fermions}
Similar to the pure compact U(1) gauge theory, also the compact QED
with staggered fermions on the lattice is known to have both the
confinement and the Coulomb phases. In the strongly coupled
confinement phase the chiral symmetry is broken in a similar way as in
QCD. The possibility to construct the continuum limit in this phase
would thus imply the existence of an abelian continuum theory with
many phenomena analogous to QCD. However, it would be strongly
interacting at short distances.

The existence of the continuum limit when approaching the phase
transition between both phases from the Coulomb phase would mean that
the usual QED can be extended to strong coupling in such a way that
a new, non-Gaussian fixed point is encountered. This might solve the
old problem of the Landau pole and of the triviality of QED, suggested
by the perturbation theory.

These considerations represent a strong impetus to investigate the
phase transition in the compact QED with staggered fermions. Earlier
investigations using the Wilson action for the gauge field found a
strong 1$^{\rm st}$ order transition, however \cite{KoDa87}. This does
not allow to perform the continuum limit, and therefore the compact
theory has been abandoned and the noncompact one prefered. Later it
has been pointed out \cite{Ok89} that, similar to the pure gauge
theory, the strength of the 1$^{\rm st}$ order transition decreases
with decreasing $\gamma$ if the extended Wilson action is used, and
that a 2$^{\rm nd}$ order transition can be found at sufficiently
large negative $\gamma$.

We have tried to check this conjecture \cite{CoFr97a}. With better
statistics than in \cite{Ok89} the appearance of the 2$^{\rm nd}$ order
could not be confirmed. However, the weakening of the two-state signal
is a fact. Of course, in all these investigations the toroidal
lattices were used. The experience with the pure gauge theory suggests
that the use of the spherical lattice might reveal a 2$^{\rm nd}$
order. Unfortunately, because of the irregularities on the lattice
constant scale, the introduction of the staggered fermions on the
spherical lattice is as yet an open problem.

Nevertheless, assuming that at $\gamma < 0$ a 2$^{\rm nd}$ order part
of the confinement-Coulomb phase transition can be found, one can
investigate its properties in the quenched approximation. For this
purpose it is sufficient to use toroidal lattices and stay at some
distance from the phase transition line at $\gamma = -0.2$, in order
to avoid the spurious two-state signal.  First results \cite{CoFr97a}
suggest that also the meson masses in the confinement phase scale
in a similar way as the pure gauge observables. They might thus survive
in the continuum limit, which might be nontrivial.

\section{Endpoint in the compact scalar QED}
A natural step in the search for critical points in the strongly
coupled compact QED is to introduce a scalar field $\phi$ of unit charge:
    \begin{eqnarray*}
     S_{U\phi} & = & S_U + S_\phi\\
           S_U & = & {\beta} \sum_P (1-\mathop{\rm Re}\nolimits {U_P})\\ 
        S_\phi & = & - {\kappa} \sum_x \sum_{\mu=1}^2
      (\phi^\dagger_x U_{x,\mu} \phi_{x+\mu} + h.c.).
    \end{eqnarray*}
The $|\phi|=1$ constraint does not restrict the model qualitatively.

As is well known, when the hopping parameter $\kappa$ increases, the
scalar field induces a phase transition. At weak gauge coupling,
$\beta > 1$, this is the Higgs phase transition between the Coulomb
phase and the Higgs region of the confinement-Higgs phase.  Its
numerical investigation in the eighties \cite{JaJe85a,JaJe86} strongly
suggested that it is of weak first order, and thus does not influence
the triviality of the Higgs sector of the standard model.

A possible candidate for the continuum limit remained to be the critical
endpoint E$_\infty$ of the Higgs phase transition line at low $\beta$. Early
analytic estimates and a numerical investigation of the
scaling along the first order phase transition line separating the
confinement and the Higgs regions \cite{AlAz93} suggested the mean field
value $\nu = 1/2$. 

We have confirmed recently this result by means of a study of the
Fisher zero in the vicinity of E$_\infty$ \cite{Fr97,FrJe97b}. Instead on
the 1$^{\rm st}$ order phase transition line we have approached this
point in the $\kappa$ direction at fixed $\beta = 0.848$. One must
take into account that the divergence of the correlation length $\xi$
is then determined by the critical exponent $\tilde{\nu}$,
\begin{equation}
  \xi\propto  (\kappa-\kappa_c)^{-\tilde{\nu}},
\end{equation}
which is different from the standard $\nu$ defined along the 1$^{\rm
  st}$ order line or its continuation. This is analogous to the
situation e.g. in the Ising model, when the critical point is
approached so that both the temperature and the external magnetic
field vary simultaneously. The equation of state predicts $\tilde{\nu}
= \nu/(\beta + \gamma)$. The mean-field values $\beta=1/2$, $\gamma=1$
and $\nu=1/2$ lead then to the prediction for a Gaussian fixed point
value $\tilde{\nu}=1/3$.

\begin{figure}[tbp]
  \begin{center}
    \leavevmode
     \psfig{file=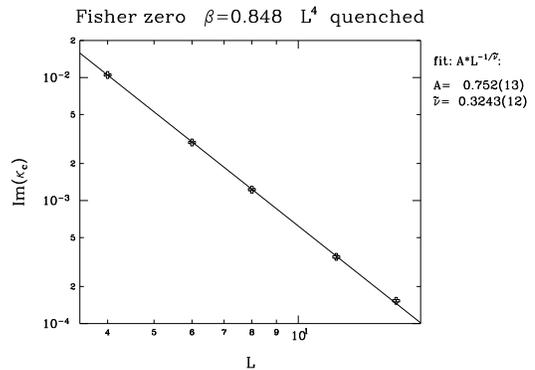,width=7.8cm,angle=90}
   \caption{FSS of $Im(z_0)$ at the endpoint in the scalar QED. }
    \label{fig:4}
  \end{center}
\end{figure}

The Fisher zero $z_0$ in the complex $\kappa$ plane scales as
\begin{equation}
  Im(z_0) = A \cdot L^{-1/\tilde{\nu}}.  
\end{equation}
The data for $Im(z_0)$ are shown in Fig.~\ref{fig:4}.  The fit gives
$\tilde{\nu} = 0.324(2)$ and is thus consistent with the mean-field
value. An analogous study in the $\beta$ direction results in
$\tilde{\nu} = 0.322(2)$. Similar results have been obtained some
years ago when the endpoint was approached in the SU(2) Higgs model
\cite{BoEv90b,Bo90}.

Thus we have confirmed by means of the FSS of the Fisher zero the
Gaussian character of the endpoint in the compact scalar QED.
Simultaneously, we have checked the applicability of the method in the
situation when the critical point is approached on some general path
in a multiparameter space. This we expect to be of much use in the
study of the tricritical point described in the next section.


\section{Compact QED with fermions and scalars of unit charge}
\subsection{Action and phase diagram}
The models discussed until now are presumably only of academic
interest for the 4D QFT. In order to create a situation more realistic
for particle physics we now introduce into the compact U(1) gauge
theory simultaneously the fermion $\chi$, and the scalar $\phi$ of unit
charges.  The main motivation is the possibility to break
spontaneously a chiral symmetry and to obtain a massive and unconfined
fermion $F=\phi^\dagger\chi$.  This resembles the Higgs-Yukawa sector
of the SM, but the symmetry breaking is of dynamical origin. The idea
is described in Ref.~\cite{FrJe95a} and the earlier results obtained
during its pursuit are presented in
Refs.~\cite{FrJe95b,FrLu95,LuFr95,FrFr95a,FrFr95b,FrJe96a}.

The action of this ``$\chi U\phi_4$ model'' is
    \begin{eqnarray*}
      S_{\chi U\phi} &=& S_\chi + S_U + S_\phi\\
      S_\chi & = & \frac{1}{2} \sum_x \overline{\chi}_x
      \sum_{\mu=1}^2 \eta_{x\mu} (U_{x,\mu} \chi_{x+\mu} - U^\dagger_{x-\mu,\mu}
      \chi_{x-\mu})\\
       & &+{am_0} \sum_x \overline{\chi}_x \chi_x\\ 
      S_U & = &{\beta} \sum_P (1-\mathop{\rm Re}\nolimits {U_P})\\ 
      S_\phi & = & - {\kappa} \sum_x \sum_{\mu=1}^2
      (\phi^\dagger_x U_{x,\mu} \phi_{x+\mu} + h.c.).
    \end{eqnarray*}
    The model has a global U(1) chiral symmetry in the limit case $m_0
    = 0$.


\begin{figure}[tbp]
  \begin{center}
    \leavevmode
     \psfig{file=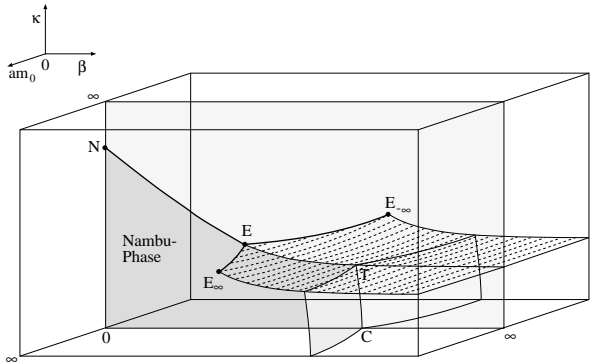,width=7.8cm,angle=0}
   \caption{Schematic phase diagram of the $\chi U\phi_4$ model. }
    \label{fig:5}
  \end{center}
\end{figure}%

The schematic phase diagram is shown in Fig.~\ref{fig:5}. We recognize the
models previously discussed as special cases of the $\chi U\phi_4$ model:
\begin{itemize}
  \item At $\kappa = 0$ and $am_0 = \infty$, the pure gauge theory with
    the Wilson action ($\gamma = 0$).
  \item At $\kappa = 0$ and $am_0$ finite, the gauge theory with fermions.
  \item At $am_0 = \infty$ and $\kappa$ arbitrary, the scalar QED.
\end{itemize}
The $\beta = 0$ case corresponds to the Nambu--Jona-Lasinio (NJL) model, as
the bosonic fields can be integrated out \cite{LeShr87a}.

At strong coupling, $\beta < 1$, the model has three sheets of 1$^{\rm
  st}$ order phase transitions: The two ``wings'', separating the
confinement and Higgs regions also at finite $am_0$, and the sheet at
$am_0 = 0$, separating the regions with nonzero chiral condensate of
opposite sign.  These three sheets have critical boundary lines
EE$_{\pm\infty}$ and NE, respectively. We have verified that within
the numerical accuracy these 2$^{\rm nd}$ order phase transition lines
indeed intersect at one point, the tricritical point E. There is no
known theoretical argument why this should be so.

\begin{figure}[tbp]
  \begin{center}
    \leavevmode
     \psfig{file=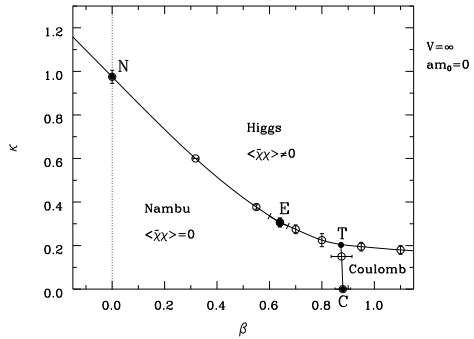,width=7.8cm,angle=90}
   \caption{Coincidence of the chiral and Higgs phase transitions on
     the line ET. }
    \label{fig:6}
  \end{center}
\end{figure}

To illustrate the existence of the tricritical point, we show in
Fig.~\ref{fig:6} the positions at small $am_0$ of the chiral phase
transition line (NETC), and of the Higgs phase transition line (ET and
its continuation to larger $\beta$). Both transitions coincide on the
ET line. There is also no cusp at E (a cusp would exclude E being a
tricritical point).


\subsection{Spectrum in the Nambu phase}
Of most interest is the Nambu phase at $m_0 = 0$, at small $\beta$ and
$\kappa$. Because of confinement there is no $\phi$-boson, i.e.
charged scalar, in the spectrum.  There is also no fundamental
$\chi$-fermion. The chiral symmetry is dynamically broken, which leads
to the presence of the composite physical fermion $F=\phi^\dagger\chi$
with the mass $am_F>0$. It scales, $am_F\searrow 0$, when the NE line
is approached.

Further states include the ``mesons'', i.e. the fermion-antifermion
bound states: the Goldstone boson $\pi$ with
$m_\pi\propto\sqrt{am_0}$, the scalar $\sigma$, and the vector $\rho$.
Also ``bosons'', the scalar-antiscalar bound states are present; the
scalar $S=\phi^\dagger\phi$ and the vector of a similar structure. Their
mixing with mesons is expected. Finally there are some gauge-balls.

\begin{figure}[tbp]
  \begin{center}
    \leavevmode
     \psfig{file=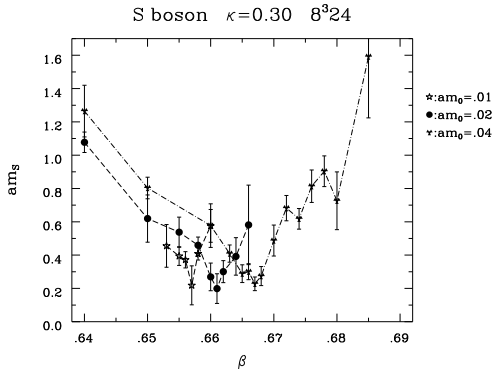,width=7.8cm,angle=90}
   \caption{Minima of the scalar boson mass $am_S$ at $\kappa = 0.3$. }
    \label{fig:7}
  \end{center}
\end{figure}

\begin{figure}[tbp]
  \begin{center}
    \leavevmode
     \psfig{file=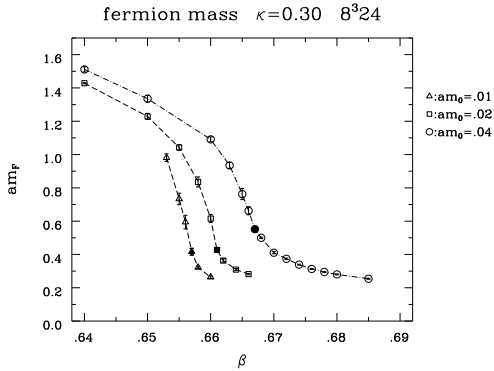,width=7.8cm,angle=90}
   \caption{Vanishing of the fermion mass $am_F$ at $\kappa = 0.3$. }
    \label{fig:8}
  \end{center}
\end{figure}

The mass $am_S$ of the $S$-boson vanishes on the lines
EE$_{\pm\infty}$, whereas $am_F$ vanishes at $am_0 = 0$ on the line NE
and above it. This provides another check of the crossing of these
three lines at the tricritical point. In Figs.~\ref{fig:7} and
\ref{fig:8} we show the minima of $am_S$ and the approximate vanishing
of $am_F$, respectively, at three values of $am_0$. Their positions
shift in $\beta$, as $am_0$ decreases, but approach the same value at
$am_0 = 0$, the approximate $\beta$ coordinate of the tricritical
point.

The spectrum is of interest mainly for an investigation of the
renormalization properties of the model, as it allows to estimate the
lines of constant physics. Of course, it gives also an information
about the spectrum in the continuum limit, if that can be constructed.

\subsection{Question of the continuum limit}
In principle, the continuum limit can be considered at any point along
the whole NE line. In the Nambu phase a massive fermion F could be
expected. The question is whether the model is renormalizable. For
this the existence of the tricritical point might be crucial.

We know that at $\beta = 0$ the model is equivalent to the
nonrenormalizable NJL model \cite{AlGo95}.  This model, properly
generalized, belongs to the universality class of the Gaussian fixed
point of a 4D Yukawa theory. Strong coupling expansion in powers of
$\beta$ suggests similar scaling properties for a finite $\beta$
interval. Indeed, it has been checked \cite{FrFr95a} that the scaling
behavior nearly on the whole NE line is very similar to that at the
point N at $\beta = 0$. However, it changes qualitatively in the
vicinity of the point E.

This raises the hope that the renormalizability properties at the
point E are different from the NJL model. This conjecture is supported
by the experience that the tricritical points belong frequently to the
universality classes different from those of the adjacent critical
lines.  

\begin{figure}[tbp]
  \begin{center}
    \leavevmode
     \psfig{file=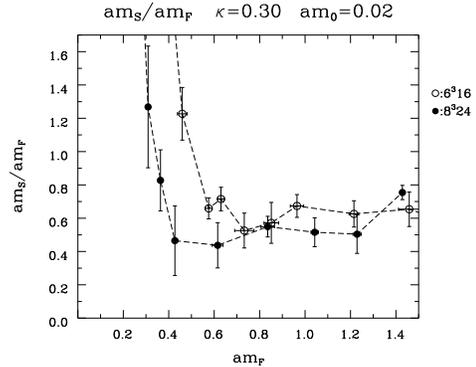,width=7.8cm,angle=90}
   \caption{ Scaling of the ratio $am_S/am_F$ at $\kappa = 0.3$. The
     turn to higher values is a finite size effect, shifting to the
     left when lattice increases.}
    \label{fig:9}
  \end{center}
\end{figure}

Furthermore, at the point E a new diverging correlation length $\xi_S
= 1/am_S$ exists, which is finite along the rest of the NE line.  In
Fig.~\ref{fig:9} we show that the ratio $am_S/am_F$ seems to approach
a constant value when the fermion mass $am_F$ decreases in the
vicinity of the point E. Thus, the continuum limit taken at the
tricritical point contains presumably at least two massive particles,
together with the massless Goldstone boson.

This, and some further results on the spectrum \cite{Fr97,FrJe97b},
suggest that the tricritical point E does not belong to the
universality class of the Gaussian fixed point of the Yukawa model
with the same symmetries, and thus merits further investigation. For
this purpose the determination of the tricritical exponent $\nu$ by
means of the Fisher zero, as in the pure gauge theory and in the
scalar QED, might be most suitable. However, with dynamical fermions
this still represents a challenge.


\acknowledgments We thank J. Cox, Ch. Hoelbling, and U.-J.~Wiese for
discussions and various informations. The computations have been
performed on the CRAY-YMP of HLRZ J\"ulich, Fujitsu VPP 500 of the
RWTH Aachen, and at the Parallel Computer Server of KFU Graz. The work
was supported by Deutsches BMBF and DFG.


\bibliographystyle{wunsnot}   


\end{document}